\documentclass[
prl
,showpacs
,showkeys
,aps
,amsfonts
,amssymb
,byrevtex
,pdftex  
,twocolumn,nofootinbib
]{revtex4-1}

\usepackage{graphicx}  
\usepackage{float}
\usepackage[titletoc]{appendix}
\usepackage{amsmath, amssymb, bm}   
\usepackage[dvips]{color}
\usepackage{natbib}
\usepackage{hyperref}

\def\){\right)} 
\def\({\left(} 
\def\]{\right]} 
\def\[{\left[}

\begin{document}

\title{
 Camouflage of the phase transition to quark matter in neutron stars}
\author{%
Wei Wei
}
\email{weiwei1981@mail.hzau.edu.cn}

\affiliation{College of Science, Huazhong Agricultural University, Wuhan, Hubei, P.R.China}
\author{%
Bryen Irving, Thomas Kl\"ahn, Prashanth Jaikumar
}

\email{thomas.klaehn@csulb.edu}

\affiliation{Department of Physics $\&$ Astronomy,
California State University Long Beach, Long Beach, CA
  90840, U.S.A. }

\begin{abstract}

It has been known for some time that compact stars containing quark matter can masquerade as neutron stars in the range of measured mass and radius, making it difficult to draw firm conclusions on the phases of matter present inside the star. Using the vector-enhanced Bag model (vBag), we examine mass-radius and mass-compactness relations with Maxwell and Gibbs construction for hybrid stars with transitions from nuclear matter to two or three-flavor quark matter, including sequential transitions. Not only can stable hybrid stars with either two or three flavor quark matter mimic neutron stars (the traditional masquerade), it appears as well difficult to distinguish two-flavor from three-flavor quark matter even in cases where a phase transition can be said to have occurred, as in the presence of a distinct kink in the mass-radius relation. Furthermore, allowing for sequential flavor transitions, we find that the transition into an unstable branch can be caused by either a transition from a nuclear to unstable quark matter or the sequential transition from nuclear to stable but ``masquerading" two-flavor to unstable three-flavor quark matter.  Addressing chiral restoration as well as quark deconfinement in a model of the phase transition, as the vBag does, adds further flexibility to the high-density equation of state, motivating caution in using even high-precision $M$-$R$ data to draw firm conclusions on the nature of phases and phase transitions in neutron stars.

\end{abstract}



\maketitle

\section{I. Introduction}
One of the reasons to study compact stars, comprising the neutron, quark or hybrid class, is to understand the state of strongly interacting matter at extreme density. Novel phases of matter are conjectured to occur inside compact stars~\cite{Blaschke:2018mqw,Endo:2013bwa,Buballa:2007rx}, but as yet there is no conclusive observational evidence of any phase that is fundamentally different from hadronic matter. Recently, the Advanced LIGO and Advanced VIRGO collaborations observed the binary neutron star merger event GW170817~\cite{GW170817}, which focused attention on using tidal polarizabilities to constrain the equation of state (EoS)~\cite{Radice,Poles,AbbottEOS,De,Chatz, Malik,Tews,Zhu,Christian:2018jyd}. It appears possible, though not conclusive, that one or both component stars in the merger could support a phase transition to quark matter at high density~\cite{Paschalidis,Nandi,Christian:2018jyd}. The so-called ``masquerade" problem~\cite{Alford04} remains: a hybrid star with quark matter in its interior cannot be easily distinguished from an ordinary neutron star based on the current observational status of masses and radii, even more so if quark matter is in a mixed phase.

In this paper, we revisit the masquerade problem, and extend it to the flavor sector of quark matter by considering phase transitions from nuclear to two-flavor (2f) or three-flavor (3f) matter, as well as a sequential appearance of the light and strange quarks that is suggested by various QCD-inspired models~\cite{Ruster,Sandin}. A recent work~\cite{Alford:2017qgh} investigated sequential transitions using the generic CSS parameterization~\cite{Alford:2013aca,Ranea-Sandoval:2015ldr} and concluded that twin or even triplet hybrid configurations can exist (a fourth family of compact stars), with similar mass but different radii. In this work, we use a specific model inspired by non-perturbative features of QCD, the vBag model~\cite{Klahn1} and find that the parameter range allows for similar mass-radius ($M$-$R$) relations for stars with phase transitions to two or three flavors.  We term this the ``flavor camouflage" problem, since it has some features different from the classical masquerade problem which is the mimicry of $M$-$R$ relations for the nuclear and mixed phase. In particular, the sequential transition from two to three flavor matter can lead to an unstable hybrid branch, irrespective of the type of construction in the crossover region (Maxwell or Gibbs). This can mimic first order transitions to two or three flavor matter that also generate an unstable branch~\cite{Alford:2013aca}. Essentially, quark matter in its various incarnations might be very effectively camouflaged in the normal neutron star branch, so other than precise observations of dramatically small radii (eg. $R\leq 9.5$km~\cite{Baus} or a radius ``gap" of about 3km~\cite{Ayriyan}), there is no smoking gun for phase transitions to quark matter in the $M$-$R$ behavior. We identify somewhat more pronounced differences in the mass-compactness curve ($M$-$C$) for the two and three-flavor case in the Maxwell/Gibbs scenario but the main conclusion is that very precise $M$-$R$ or $M$-$C$ data is still required to resolve either the masquerade or the flavor camouflage problem. Certainly, given the current state of knowledge of dense quark matter from effective theories, it would be premature to conclude even from high-precision mass and radius numbers that are neutron-star like, whether a phase transition is supported or not.

 This paper is organized as follows: section II describes the nuclear and quark matter equations of state we apply, where we put explicit focus on the vBag EoS, as the conclusions of this paper are based on the variable range of those EoS parameters. Section III serves to briefly review Maxwell- and Gibbs-type phase transitions, that is, first order phase transitions in neutron star matter under local and global conservation of electric charge. In Section IV we discuss a number of different scenarios for the mass-radius relation of hybrid stars with a summary of our conclusions in Section V.
 
\section{II. Nuclear and Quark Matter Model Equations of State}
\subsection{Nuclear Matter}

A common approximation for the energy per particle in bulk matter is given by the parabolic expansion
\begin{equation}
    E(n,x)=E(n,x=\frac{1}{2})+S(n)(1-2x)^2,
\end{equation}
where $x$ is the proton fraction, $E(n,x=\frac{1}{2})$ is the energy per particle of symmetric nuclear matter, and $S(n)$ is the nuclear symmetry energy. For the purpose of our paper - to study possibilities to camouflage hybrid stars - the specific choice of
a nuclear equation of state is not our main concern.
We apply the SL model ~\cite{PAL,PBP} for which the total energy per particle of symmetric nuclear matter is defined as
\begin{equation}
    E(n,x=\frac{1}{2})=\frac{3}{5}E_F^{(0)}n_0u^{\frac{5}{3}}+V(u),
\end{equation}
where $E_F^{(0)}$ is the Fermi energy of nuclear matter at the equilibrium density and $u=n/n_0$ with $n_0$ denoting equilibrium nuclear matter density. The potential contribution is parameterized as 
\begin{equation}
   V(u)=\frac{1}{2}An_0u^2+\frac{Bn_0u^{\delta+1}}{1+B'u^{\delta-1}}+n_0u^2\sum_{i=1,2}C_i[1-\frac{3}{5}\frac{u^{2/3}}{R_i^2}],
\end{equation}
where $R_i=\Lambda_i/\hbar k_F^{(0)}$, $k_F^{(0)}$ is the Fermi momentum of nuclear matter at saturation and $\Lambda_i$ is a finite-range force parameter. To separate the kinetic and potential contributions to
the symmetry energy, we write
\begin{equation}
    S(n)=(2^{2/3}-1)\frac{3}{5}E_F^{(0)}[u^{2/3}-F(u)]+S_0F(u),
\end{equation}
where $S_0=30$ MeV and $F(u)$ defines the potential contributions to the symmetry energy and $F(1)=1$. Table 1 offers three parameterizations where the potential part of the symmetry energy varies approximately as $\sqrt{u}$, $u$ and $2u^2/(1+u)$, respectively.
The parameters $A$, $B$, $\delta$, $C_1$, $C_2$, and $B'$ are determined from constraints provided by the empirical properties of symmetric nuclear matter at saturation density $n_0 = 0.16 fm^{-3}$.
For this work SL23 was chosen as the nuclear EoS.
\begin{table}[]
    \centering
    \begin{tabular}{c|c|c|c|c|c|c|c}
    \hline\hline
    EoS  & $K_0$ & A & B & B' & $\delta$ & $C_1$ & $C_2$ \\
     \hline
    SL13 & 120 & 3.706 & -31.155 &  0 & 0.453 & -41.28 & 43\\
    SL23 & 180 & 159.47 & -109.04 &  0 & 0.844 & -41.28 & 43 \\
    SL33 & 240 & 204.01 & 72.704 &  0.3 & 1.235 & -41.28 & 43 \\
    \hline
    \end{tabular}
    \caption{The notations $SLn_1n_2$ is used
to denote different EoS; $n_1=1,2,3$ indexes increasing values of the incompressibility $K_0$, and $n_2 = 1,2,3$ indicate, respectively, a  $\sqrt{u}$, $u$ and $2u^2/(1+u)$ dependence of the nuclear symmetry potential energy $F(u)$ on the density (see text above).}
    \label{tab:my_label}
\end{table}

\subsection{Quark Matter - Extended Bag Model with Vector Interactions}
None of the currently available QM EoS is obtained from first principle QCD-based calculations.
QCD is inherently complicated and this holds even more under the conditions in a compact star's core : large chemical potentials $\mu_B>m_n$ and comparably small temperatures, $T/\mu_B\approx 0$.
A number of quark matter model EoS have been developed to account for key phenomena of QCD which usually means
deconfinement and the restoration of chiral symmetry at high density. While the effect of chiral symmetry restoration is tied to the dressed quark mass which approaches the bare quark mass at
sufficiently high density, the actual mechanism behind confinement and deconfinement in dense and cold nuclear matter is not well-understood.
There are several models of confinement applied to compact star structure~\cite{Blaschke:2010vj,Contrera:2012wj} but they are not based on an unambiguous order parameter for deconfinement.
A consistent equation of state for the two phases, nuclear and quark matter, that is derived from QCD, does not exist and so the standard approach is to have separate phenomenological descriptions of the two. Typically, one would account for confined and deconfined matter with two different model EoS, naturally understood as the nuclear EoS and quark matter EoS.
For the quark matter branch, chiral symmetry restoration is either dynamically modeled or quarks enter the model as bare quarks. Prominent examples of the first class are all kinds of Nambu-Jona-Lasinio (NJL) models where contact-interaction terms in the scalar channel
result in chiral symmetry breaking approximately at quark chemical potentials smaller than the dressed quark mass in vacuum. The thermodynamic Bag model stands for the second type of EoS and is a limiting case of the well known MIT-Bag model which had been developed to describe confined hadrons.

For the purpose of our study we will use the vBag model \cite{Klahn1}.
It has been introduced to consolidate a number of seemingly discrepancies between the NJL and Bag model.
In the thermodynamic bag model, pressure and energy density of quark matter are given by free Fermi gas expressions
and an additional phenomenological bag constant (which has been originally adjusted to reproduce hadron masses),
\begin{eqnarray}
P_{\rm Bag}&=&\sum_{f=u,d,s}P_{FG,f}(\mu_f,m_f)-B\\
\varepsilon_{\rm Bag}&=&\sum_{f=u,d,s}\varepsilon_{FG,f}(\mu_f,m_f)+B
\end{eqnarray}
Unlike for NJL models, there is no chiral symmetry breaking; $m_f$ is the bare mass of each individual flavor.
Subtracting the positive definite bag constant implies a threshold chemical potential for quark matter
at which the pressure turns positive and therefore gives a physically meaningful description of deconfined quark matter.
It should be noted that $B$ is flavor blind, viz. the model activates all light flavors 
including the strange quark at once if $B$ is sufficiently large.
For NJL models, contact interactions in the scalar channel result in positive pressure as soon as chiral symmetry restoration sets
in. The pressure in the chirally broken phase is zero. While this behavior seems similar in the chirally restored
phase, there is one major difference, which is the flavor dependence of the effect, viz. dependent on the
bare quark mass, chiral symmetry is restored at different quark chemical potentials for each individual quark flavor.
In \cite{Klahn1} it has been illustrated that the ideal gas approximation in combination with a bag constant
is still a reasonable approximation of NJL model results in the chirally restored phase.
To ensure agreement with NJL model results, one introduces a chiral bag constant for each flavor independently. 
The existence of a Bag like term in NJL model EoS is related to the existence of a scalar vacuum condensate;
the melting of this condensate is the key mechanism to describe chiral symmetry breaking.

A further difference between the thermodynamic Bag and NJL type models is the lack of vector repulsion in the thermodynamic Bag model.
This results in a rather soft equation of state which consequently could not account for the existence
of massive neutron stars with quark matter core. 
However, this problem can be overcome by accounting for perturbative $\alpha_s$ corrections which reduce
the pressure at given chemical potential. 
As noted in \cite{Farhi:1984qu}, the size of the perturbation parameter is not small enough
to safely assume correctness of this approach which after all is not that surprising in a transition domain
which is expected to be strongly influenced by non-perturbative effects.
In NJL type models, scalar and vector interactions both appear on an equal footing, which
can be understood if one traces these interactions back to the underlying quark-gluon interaction.
The vBag model is a hybrid approach which accounts for scalar interactions and hence chiral symmetry breaking
in a slightly simplified, bag model like, way, viz. by assuming bare quark masses and flavor dependent 
chiral bag constants to reproduce the proper critical chemical potential for each flavor's chiral transition. 
Vector interactions are taken into account non-perturbatively and in full analogy to the NJL model.
This results in the following expression for the pressure of a single flavor:
\begin{eqnarray}
P_{{\rm vBag},f}&=&P_{FG,f}(\mu_f^*)+\frac{K_v}{2}n_{FG,f}^2(\mu_f^*)-B_{\chi,f}\quad,
\end{eqnarray}
where the second term results from vector interactions at given coupling constant $K_v$.
As in the NJL model, the effective flavor chemical potential $\mu_f^*$ has to be determined
self-consistently at given bare flavor chemical potential $\mu_f^*$,
\begin{eqnarray}
\mu_f=\mu_f^*+K_v n_{FG,f}(\mu_f^*).
\end{eqnarray}
At large densities, all flavors are activated. 
One would assume that $B_\chi=\sum_f B_{\chi,f}$ would reproduce the bag model bag constant.
However, the NJL model is known to predict a larger bag constant than the bag model.
This problem resolves itself if one recalls that the bag model is adjusted to reproduce the masses of hadrons, that is confined objects, while the original NJL model approach has no built-in confinement.
If confined quarks are energetically favorable, the bound state energy should
be lowered and hence the effective bag constant smaller than the bag constant of
a model without inherent confinement.
vBag accounts for this reduction by introducing a confinement bag constant $B_{\rm dc}$ which 
is added to the total pressure and consequently subtracted from the total energy density,
\begin{equation}
P_{{\rm vBag}}=\sum_f P_{{\rm vBag},f}+B_{\rm dc}\quad.
\end{equation}
Interestingly, in this way, the vBag approach to account for chiral symmetry restoration
elucidates that the MIT bag model's ``bag"
represents to a large extent the scalar condensate while a confining background 
term reduces the energy content,
\begin{equation}
B_{\rm MIT}=\sum_f B_{\chi,f} - B_{\rm dc}.
\end{equation}
For the purpose of this work we introduce a two-flavor and a three-flavor
version of vBag, where
\begin{equation}
P_{{\rm vBag}}^{2f}=\sum_{f=u,d} P_{{\rm vBag},f}+B_{\rm dc}\quad,
\end{equation}
and
\begin{equation}
P_{{\rm vBag}}^{3f}=P_{{\rm vBag}}^{2f} + P_{{\rm vBag},s}.
\end{equation}
Underlying this notation is the idea that the
dressed quark mass of strange quarks is larger than that of the
lighter up and down quarks and therefore the chiral transition
can take place at a larger chemical potential, specified by the condition
\begin{equation}
P_{{\rm vBag},s}(\mu_{s,\rm crit}^*)=0,
\end{equation}
as explained earlier.
The most extreme scenario we consider is the bag model limit, 
$B_{\chi,s}=0$ which allows for the instant appearance of three-flavor matter 
under the condition that $P_{{\rm vBag}}^{2f}=0$ at non-zero chiral bag constant
and deconfinement bag constant for light quarks.
Since we perform a parameter study which looks for qualitative effects rather
than final quantitative results, we will further refer to an effective two flavor bag
constant,
\begin{equation}
B_{\rm eff}^{2f}=B_{\chi,u}+B_{\chi,d}-B_{\rm dc}.
\end{equation}
Although the chiral bag constant fixes
the chiral transition density in two flavor matter, adding the deconfinement bag constant
ensures that the actual phase transition happens at the same chemical potential.
For this parametric study only the total effect matters, which is
quantified by $B_{\rm eff}^{2f}$.

\section{III. Phase Transitions}
Before we apply the model to hybrid stars, let us review the standard lore of phase transitions in the context of compact stars. Matter in neutron stars is charge neutral and $\beta$-equilibrated.
The latter, at low temperatures, relates the chemical potentials in
nuclear and quark matter as
\begin{eqnarray}
\mu_n&=&\mu_p+\mu_e,\\
\mu_d&=&\mu_u+\mu_e,\\
\mu_s&=&\mu_d.
\end{eqnarray}
Together with the charge neutrality conditions in each phase,
\begin{equation}
\sum_{i=n,p,e}Q_in_i=0 \quad \mbox{and} \sum_{i=u,d,s,e}Q_in_i=0
\end{equation}
it is evident that one chemical potential is sufficient to
characterize the individual thermodynamic state of nuclear and quark matter
if no phase transition is assumed.
To compare both phases one chooses the baryochemical potential $\mu_B$
which relates to the conserved baryon number and reads as
\begin{equation}
\mu_B=\mu_n \quad \mbox{and} \quad \mu_B=2\mu_d+\mu_u
\end{equation}
respectively.
If both phases are independent the phase with the higher pressure minimizes
the thermodynamical potential and is therefore energetically favorable.
Consequently, a phase transition will occur if nuclear and quark matter have
equal pressure at equal baryochemical potential,
\begin{equation}
P^H(\mu_{B,\rm crit})=P^Q(\mu_{B,\rm crit}).
\end{equation}
At $\mu_{B,\rm crit}$ both pressures have different slopes
and, as the transition follows the higher pressure, the baryon density
will show a discontinuity, typical for first order Maxwell phase transitions.

However, this picture of ``one phase or the other" is not the only possibility. 
As an alternative, one can assume that both phases can mix.
If this is the case the equilibrium conditions remain but charge neutrality
holds over both phases.
Since it is not ``one or the other" the available volume has to be split between the
two phases with each phase contributing with corresponding fraction $\eta=V_Q/(V_Q+V_H)$ and $1-\eta=V_H/(V_Q+V_H)$ 
respectively, to the total charge density,
\begin{equation}
(1-\eta)\sum_{i=n,p,e}Q_in_i+\eta \sum_{i=u,d,s,e}Q_in_i=0.
\end{equation}
Note that the electron appears in both phases and consequently contributes fully and independently of the volume fraction, $(1-\eta)n_e+\eta n_e=n_e$.
Similarly as before, the transition takes place if both phases have the same pressure
\begin{equation}
P^H(\mu_{B,\rm crit})=P^Q(\mu_{B,\rm crit})\,.
\end{equation}
However, all other quantities contribute according to their volume fraction,
e.g., the total energy density reads
\begin{equation}
\varepsilon = (1-\eta)\varepsilon^H + \eta\varepsilon^Q. 
\end{equation}
This scenario for a phase transition is well known and often referred to
as the ``Gibbs"- or ``Glendenning"-construction.
The physical manifestation of this approach looks quite different to
the Maxwell transition. There is a transition regime $\left[ \mu_{B,\rm crit}^{I}, \mu_{B,\rm crit}^{II} \right]$
in which the quark volume fraction will grow from zero to one, and due to the smooth transition
the sudden jump of density that is typical for the Maxwell transition at the critical pressure
is now replaced by a different behavior as the density increases smoothly and a discontinuity is observed only for higher derivative terms.

\section{IV. Results}
We now present the consequences of our model for the structure of compact stars. The idea that quark stars can have a mass radius relationship very similar to that of pure neutron stars is not that new and often referred to as the masquerade problem following the work in \cite{Alford04}.
Essentially any hybrid EoS constructed from two model EoS with similar slope in the transition domain
will have a small latent heat and show a rather smooth transition
from one phase into the other. This does not only hold for the transition from
nuclear to quark matter but can affect other exotic states of matter in
the same way. For instance, it is not clear to what extend the core of
a massive neutron star carries hyperons. 
We do not aim to answer this question but want to mention that if hyperons appear they do so smoothly
and would not show a distinct impact on the mass radius relationship in
form of a ``kink" or other discontinuous behavior.

For this work, our nuclear model EoS serves as a generic nuclear EoS and could be replaced by other model EoS for nuclear or hyperonic matter
without altering the main results of this work which we illustrate
in the following.

\subsection{Nuclear Matter Masquerade \\of 2f and 3f Quark Matter}

\begin{figure}[htb]
\centering
\vspace{-8mm}
\includegraphics[width=6.05cm,angle=270]{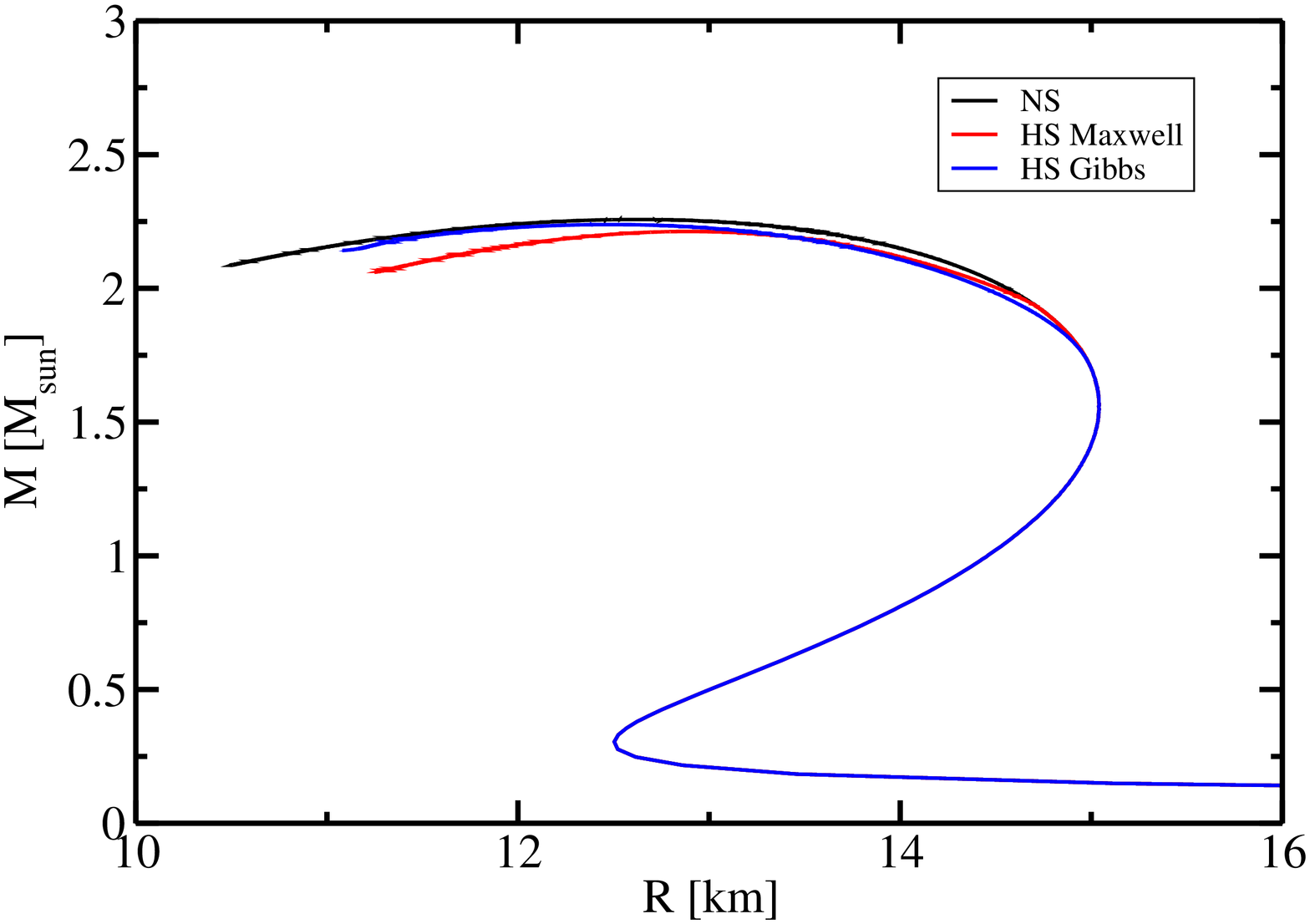}
\vspace{-3mm}
\includegraphics[width=6.05cm,angle=270]{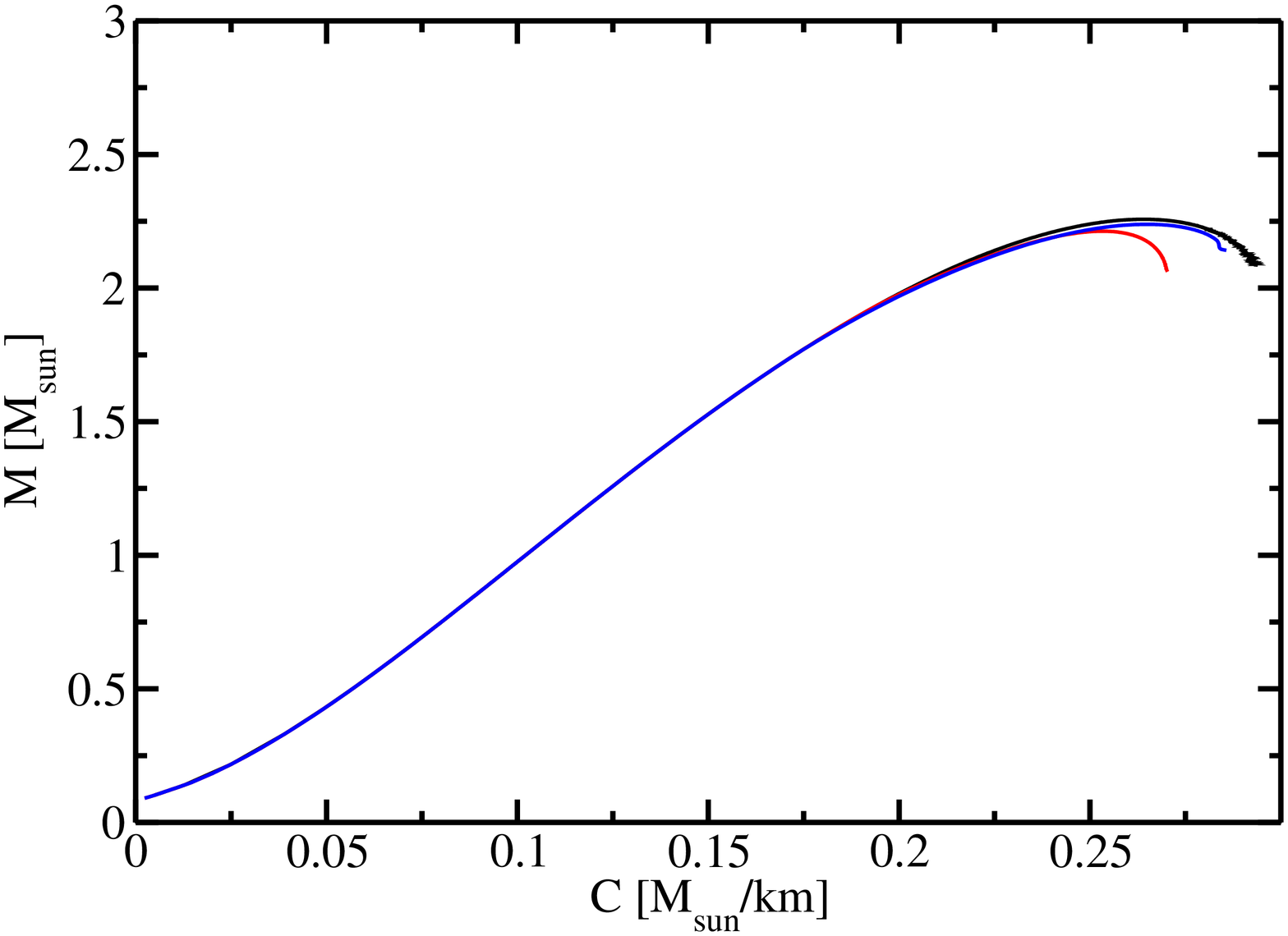}
\vspace{-5mm}
\includegraphics[width=6.05cm,angle=270]{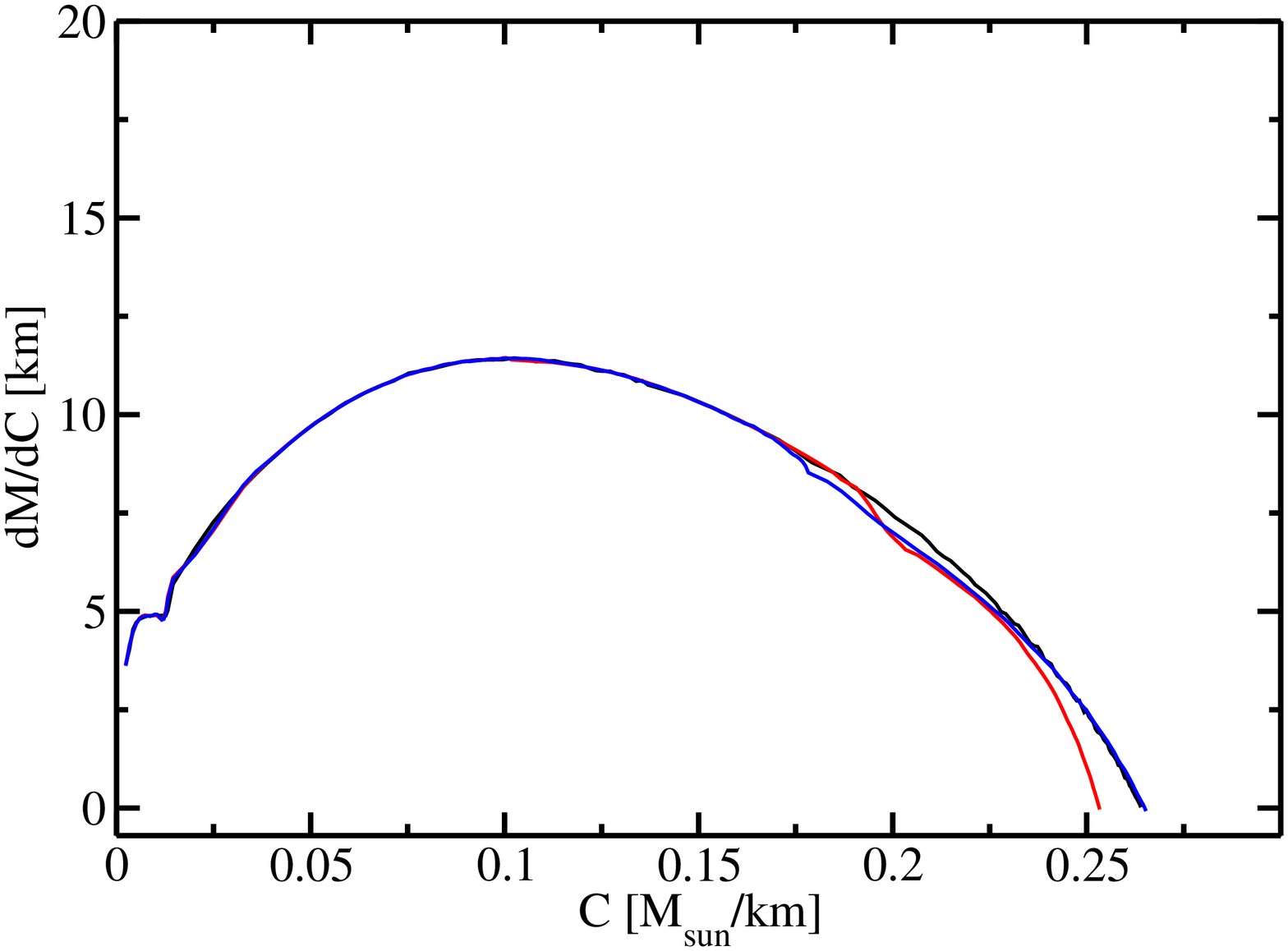}
\caption{
Nuclear masquerade of 2-flavor quark matter. The mass-radius relation of the nuclear EoS (black) can be nearly reproduced assuming a phase transition to 2f-quark matter of either Maxwell-type 
(red; $K_v=7\times10^{-6}\,{\rm MeV}^{-2}$, $B_{\rm eff}=70\, {\rm MeV\,fm^{-3}}$) 
or Gibbs-type 
(blue; $K_v=9\times10^{-6}\,{\rm MeV}^{-2}$, $B_{\rm eff}=70\, {\rm MeV\,fm^{-3}}$).
Visible variations occur only around masses of two solar masses and more.
}
\label{fig:nmasquerade2f}
\end{figure}

As discussed in \cite{Alford04} and subsequent works, we note the possibility that hybrid stars can reach and exceed a mass of two solar masses and that the
$M$-$R$ relationship may show no evident sign of a phase transition.
We want to emphasize that despite the similarity to the thermodynamical bag model,
the necessary stiffness of the vBag EoS does not arise from perturbative corrections
but solely from the non-perturbative treatment of the quark-gluon interaction in the
vector channel.
The correction terms due to vector interaction are proportional to the square of the particle number density,
which is qualitatively similar to the results one obtains in relativistic mean field models for nuclear matter.
It is therefore not too surprising that one can find parameterizations which generate quark matter EoS which
are nearly indistinguishable from purely nuclear EoS and consequently have very similar mass radius relationships.
We illustrate this in Fig.\ref{fig:nmasquerade2f}, where we chose vBag parameterizations which result in mass radius relationships nearly reproducing the purely nuclear EoS results (top panel) up to and beyond two solar masses. 
As the main purpose of this plot is to provide an example for a nearly perfect masquerade,
we chose different parameterizations for Maxwell and Gibbs construction
(otherwise the Gibbs construction would result in an earlier onset of the mixed phase and 
can look quite different from the hybrid EoS with Maxwell construction).
Visible differences between the three scenarios - nuclear, 2f-Maxwell, 2f-Gibbs - occur 
at masses beyond 2 $M_{\rm sun}$. 
There is no distinct change in the appearance of
the $M$-$R$ curve which would visibly indicate a phase transition.
For a more detailed illustration we plot the mass as a function of compactness $C=M/R$
in order to 'uncurl' the $M$-$R$ plot (middle panel).
Notice that a similar effect could be achieved by plotting the neutron star mass as function of the gravitational red-shift $z$ instead.
Unlike for $M$-$R$ curves the $M$-$C$ plot apparently provides a bijective mapping up to the maximum mass 
and therefore enables us to obtain a smooth derivative of the mass with respect to compactness in order to look for
more subtle changes (bottom panel).
The Maxwell construction leaves an imprint slightly below $C=0.2 \,\rm M_{\rm sun}/km$.

Nevertheless this imprint is small and it would require very precise $M$-$R$ measurements in order to
resolve it. In case of the Gibbs transition we notice a visible difference to the
purely nuclear case. It seems extremely hard to identify
a clear signal for a possible phase transition from the behavior of these curves.
Keeping in mind that these three curves have been obtained under very different assumptions this might be the most disturbing result of our analysis: There is no guarantee, that any seemingly smooth $M$-$R$ or $M$-$C$ data and even smooth derivatives of the latter necessarily imply a purely nuclear composition of massive neutron stars. It has to be added, that a hybrid scenario as described here would be a very unfortunate and maybe unlikely outcome. We had to fine-tune parameters (but not unreasonably so) to provide a transition smooth enough to leave that little of an imprint in the $M$-$C$ derivative.

\begin{figure}[H]
    \centering
\vspace{-8mm}
\includegraphics[width=6.05cm,angle=270]{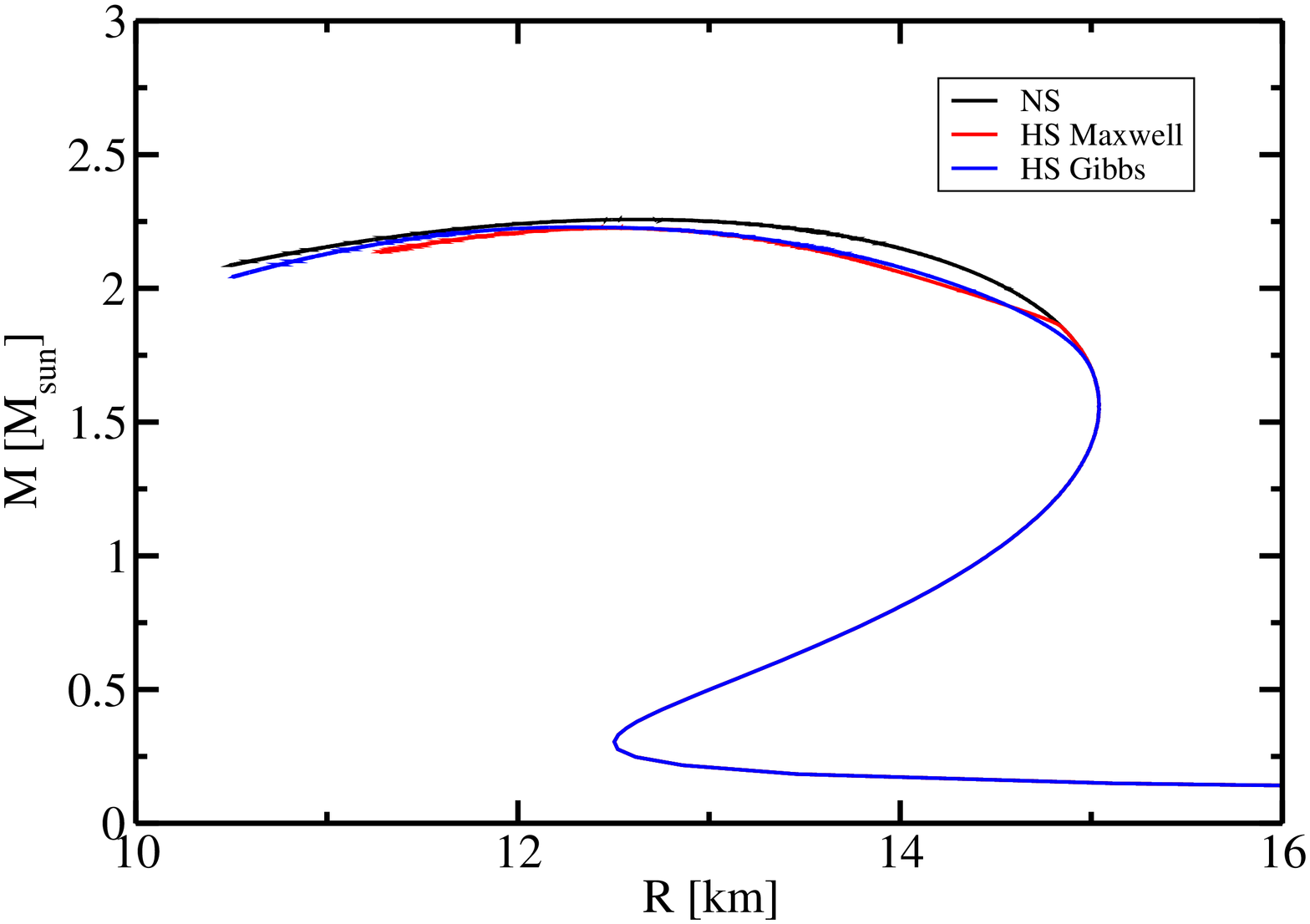}
\vspace{-3mm}
\includegraphics[width=6.05cm,angle=270]{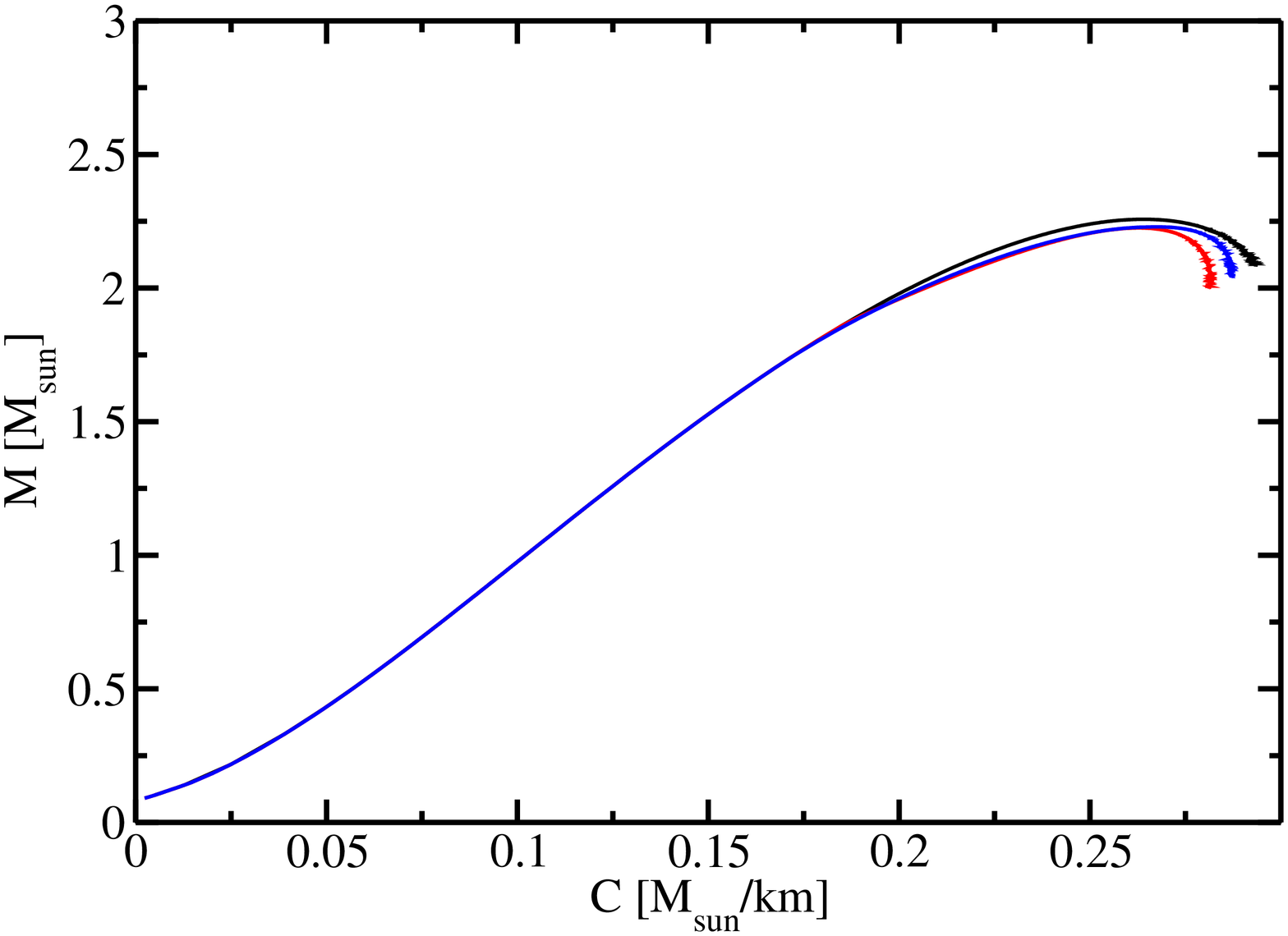}
\vspace{-5mm}
\includegraphics[width=6.05cm,angle=270]{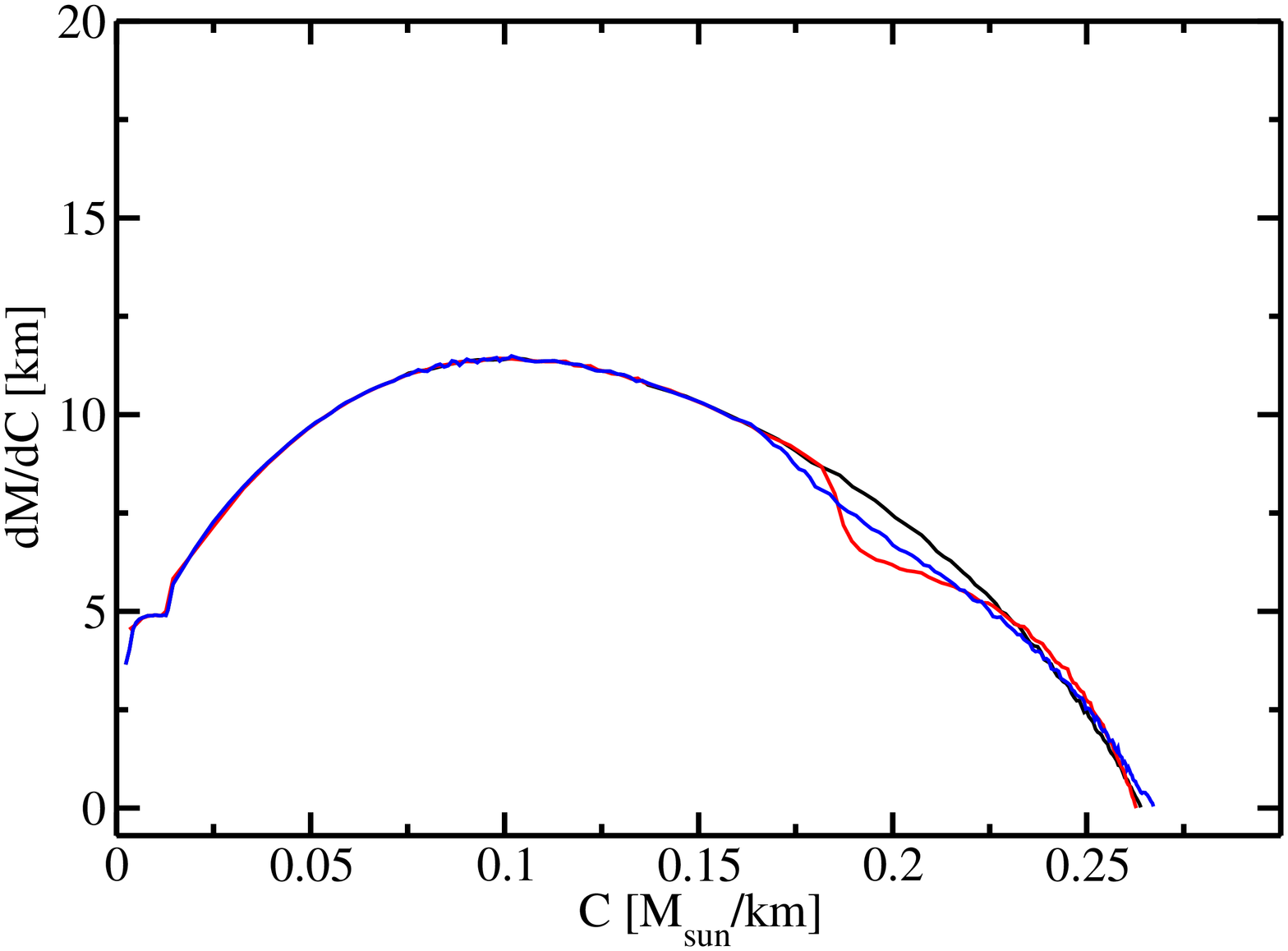}
\caption{
Nuclear masquerade of 3-flavor quark matter. The mass-radius relation of the nuclear EoS (black) can be nearly reproduced assuming a phase transition to 3f-quark matter of either Maxwell-type 
(red; $K_v=23.5\times10^{-6}\,{\rm MeV}^{-2}$, $B_{\rm eff}=78.5\, {\rm MeV\,fm^{-3}}$) 
or Gibbs-type 
(blue; $K_v=32\times10^{-6}\,{\rm MeV}^{-2}$, $B_{\rm eff}=80\, {\rm MeV\,fm^{-3}}$).
Visible variations occur only around masses of two solar masses and more.
However, the compactness derivative of the mass dM/dC reflects a sharp kink 
of the Maxwell-type and smooth transition of the Gibbs-type transition.
}
\label{fig:nmasquerade3f}
\end{figure}

In Fig.\ref{fig:nmasquerade3f} we show an only slightly less pessimistic result, this time
for a transition from nuclear to three-flavor matter. To compensate for the softening
of the quark matter EoS due to the additional strange quark degree of freedom we increased
the vector coupling significantly. 
The $M$-$R$ plot for the purely nuclear scenario seems quite distinct to the two hybrid models. 
However, the aim of this study is not to perfectly mimic a given nuclear EoS but
to assess whether we can identify a phase transition from a single $M$-$R$ relation if that data were available.
For the Maxwell case the transition results in a perceptible but tiny kink at about 1.9 $\rm M_{\rm sun}$.
For the Gibbs case this kink is smoothed out (both top panel). 
Uncurling into a mass-compactness plot (middle panel) seems to make it harder to identify the phase transition region, but
taking the derivative (bottom panel) reveals a distinct discontinuity in the Maxwell case and
a dip in the otherwise perfectly convex curve in the Gibbs case.
In conclusion, 'weak' phase transitions with only minor visible affect on the $M$-$R$ relation 
can leave a noticeable imprint on the $M$-$C$ derivative.

However useful it would be to answer the question whether a phase transition is taking place,
based on the $M$-$C$ derivative plot and even in cases where the $M$-$R$ relationship does not 
show apparent bends or kinks, this is still not enough information to decide safely 
whether the transition went into two- or three-flavor matter.
Unfortunately, the modeling of the quark matter EoS is still based on effective models,
and that being the case, it limits the inferences that can be drawn about phase transitions in the density domain of neutron stars.

\begin{figure}[H]
    \centering
\includegraphics[width=6.5cm,angle=270]{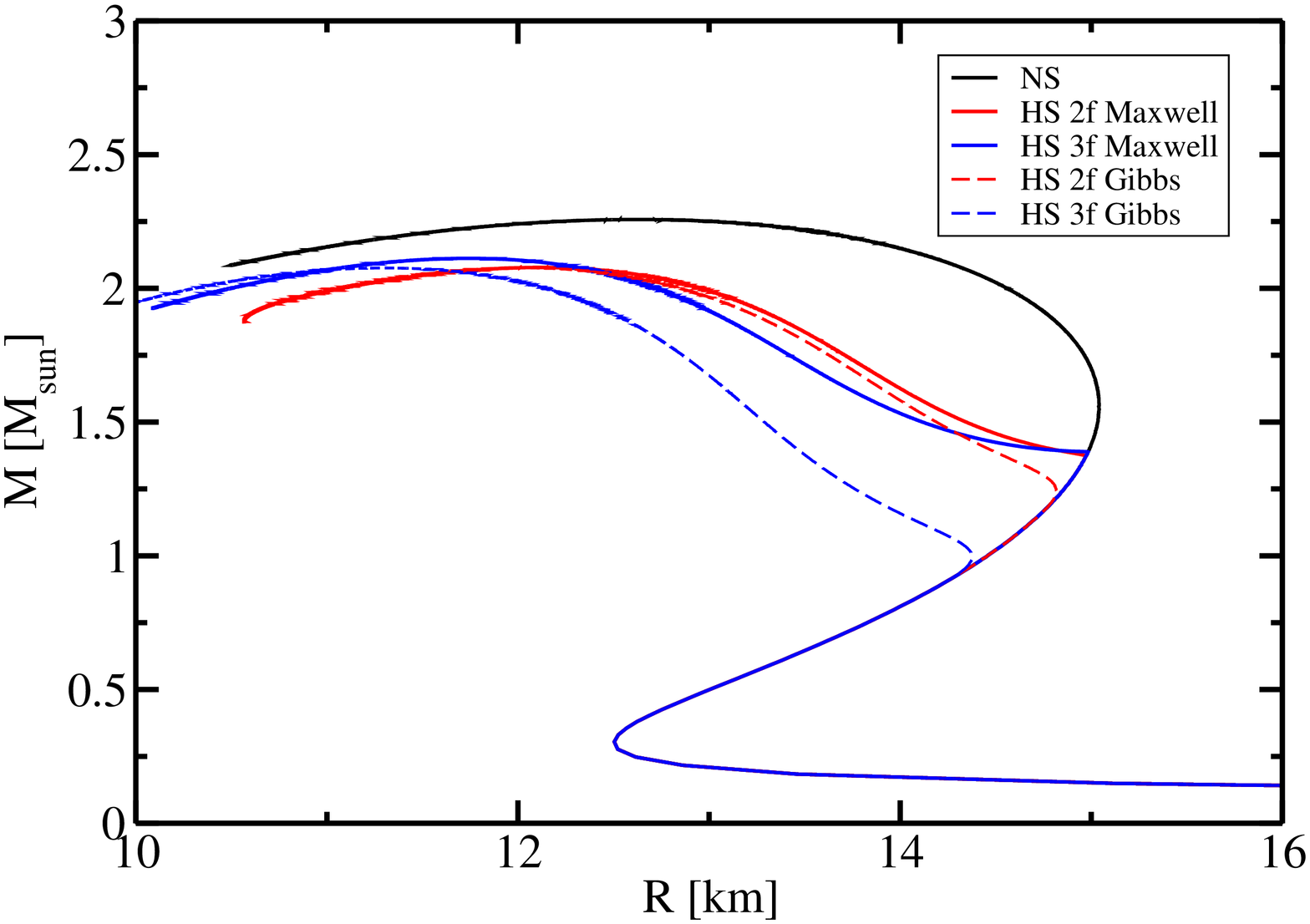}
\includegraphics[width=6.5cm,angle=270]{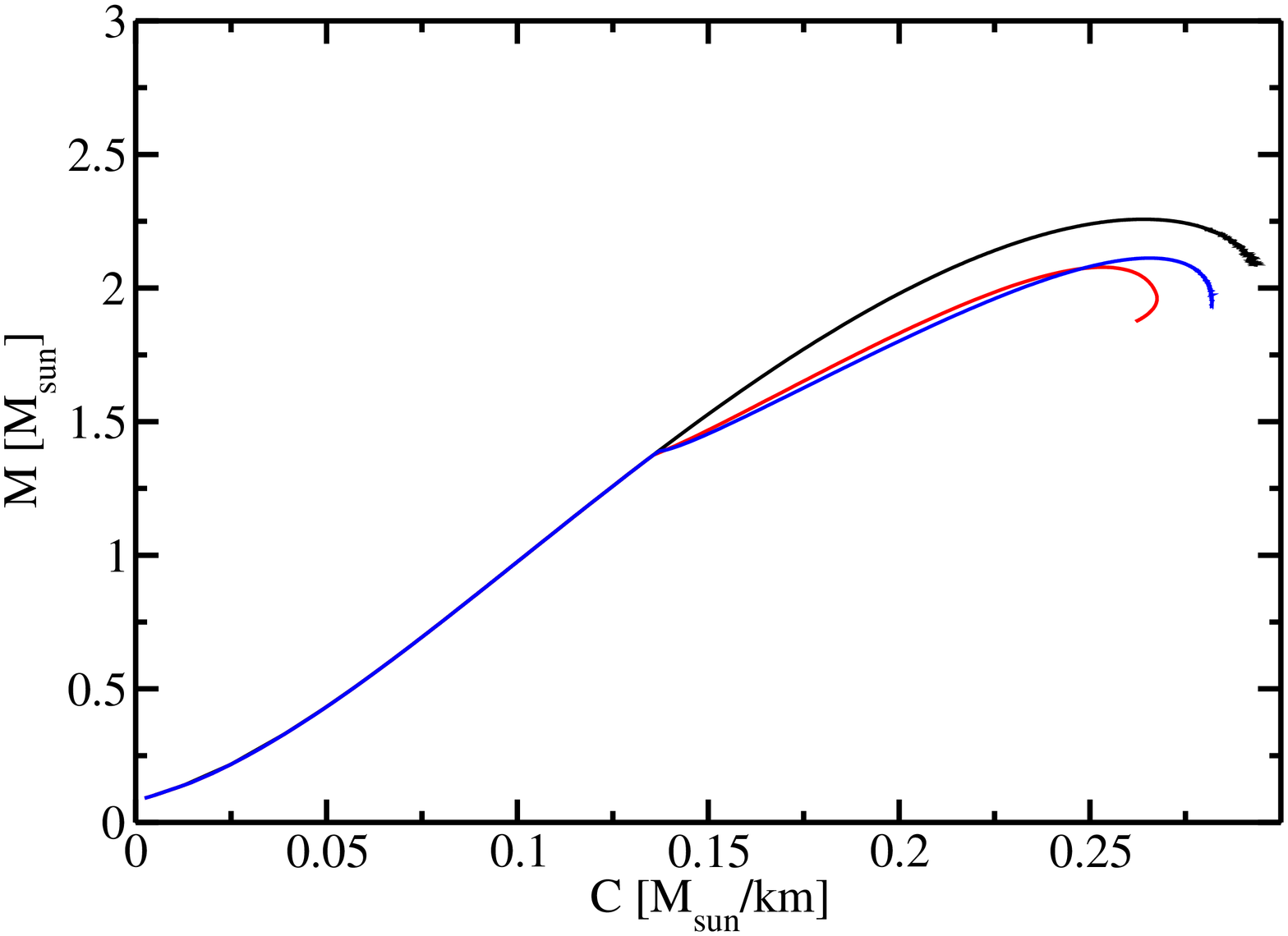}
\includegraphics[width=6.5cm,angle=270]{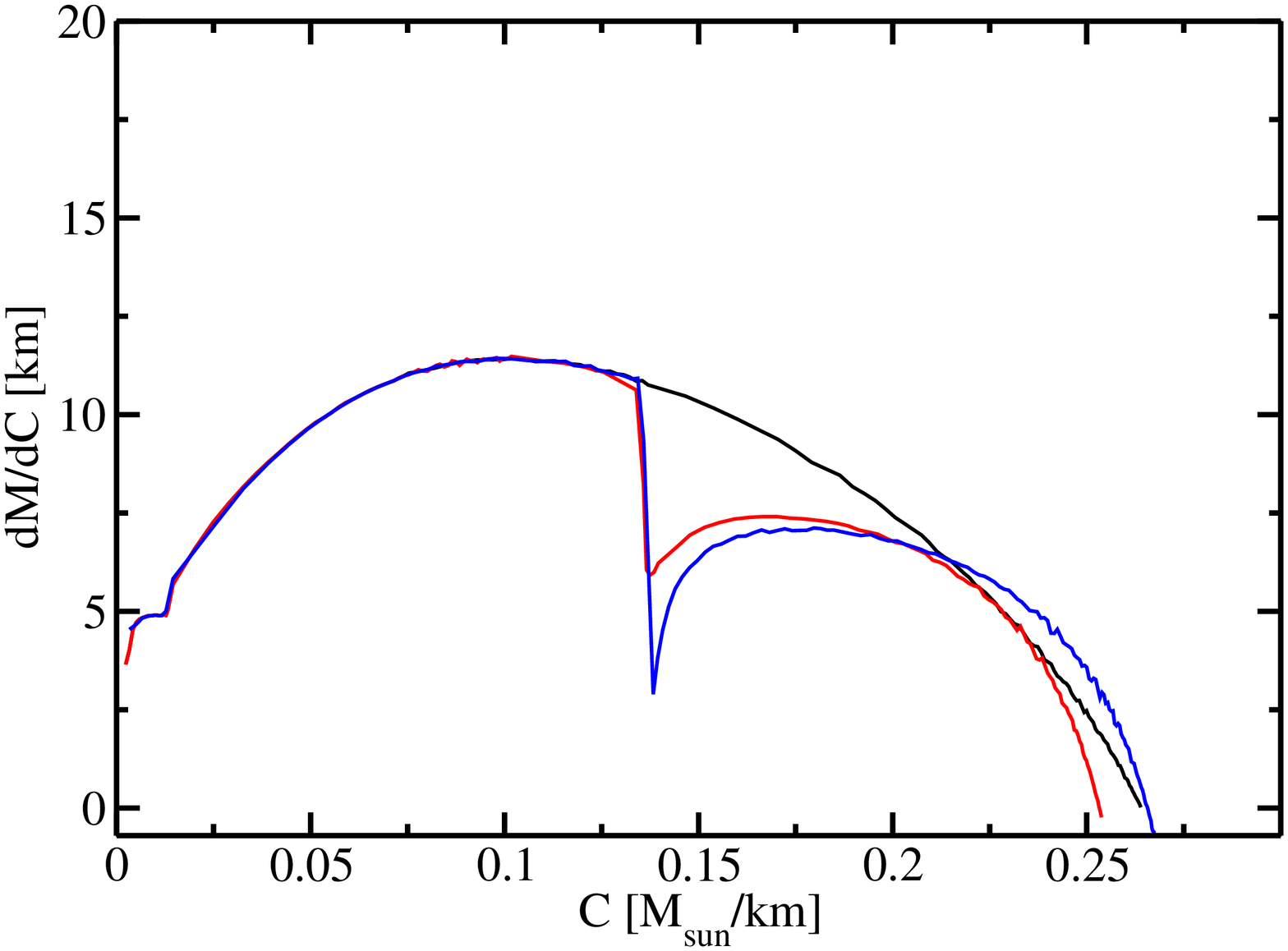}
\caption{Flavor camouflage for Maxwell type transitions: although different vBag parameterizations are introduced, the transition to 2f quark matter
(solid red; $K_v$=$5\times10^{-6}\,{\rm MeV}^{-2}$, $B_{\rm eff}$=$70\, {\rm MeV\,fm^{-3}}$)
and 3f quark matter
(solid blue; $K_v$=$20\times10^{-6}\,{\rm MeV}^{-2}$, $B_{\rm eff}$=$78\, {\rm MeV\,fm^{-3}}$) behaves very similarly in mass, radius, compactness and even compactness derivative d$M$/d$C$.
For completeness, the Gibbs type transition is also plotted as dashed lines with identical color code.
}
    \label{fig:MaxCamou}
\end{figure}

In Fig.\ref{fig:MaxCamou} we chose different parameterizations for the 2f- and 3f-case, which result in
similar $M$-$R$ curves assuming a Maxwell transition. For completeness, we provide the result one would
obtain with a Gibbs construction (all upper panel).
As before, we provide a $M$-$C$ and a $M$-$C$-derivative plot.
Although there are visible differences, the generic behavior of 2f and 3f hybrid stars is too similar
to conclude safely from data which case has been observed.

\subsection{2f-3f Camouflage}

In the previous subsection we dealt with the classical masquerade problem, viz. $M$-$R$ configurations
which do not clearly indicate a phase transition although it takes place.
We pointed out that this problem can be identified for a transition into 2f or 3f quark matter.
This possible ambiguity of quark matter EoS with different flavor degrees of freedom
we refer to as ``flavor camouflage".
In the following, we provide a few examples of cases where one {\it can}
conclude that a phase transition took place but would not be able to 
identify the flavor content of the hybrid star. 
This results from the possibility for a {\it sequential} transition
from nuclear to 2f to 3f quark matter.

\begin{figure}[H]
    \centering
\includegraphics[width=6.5cm,angle=270]{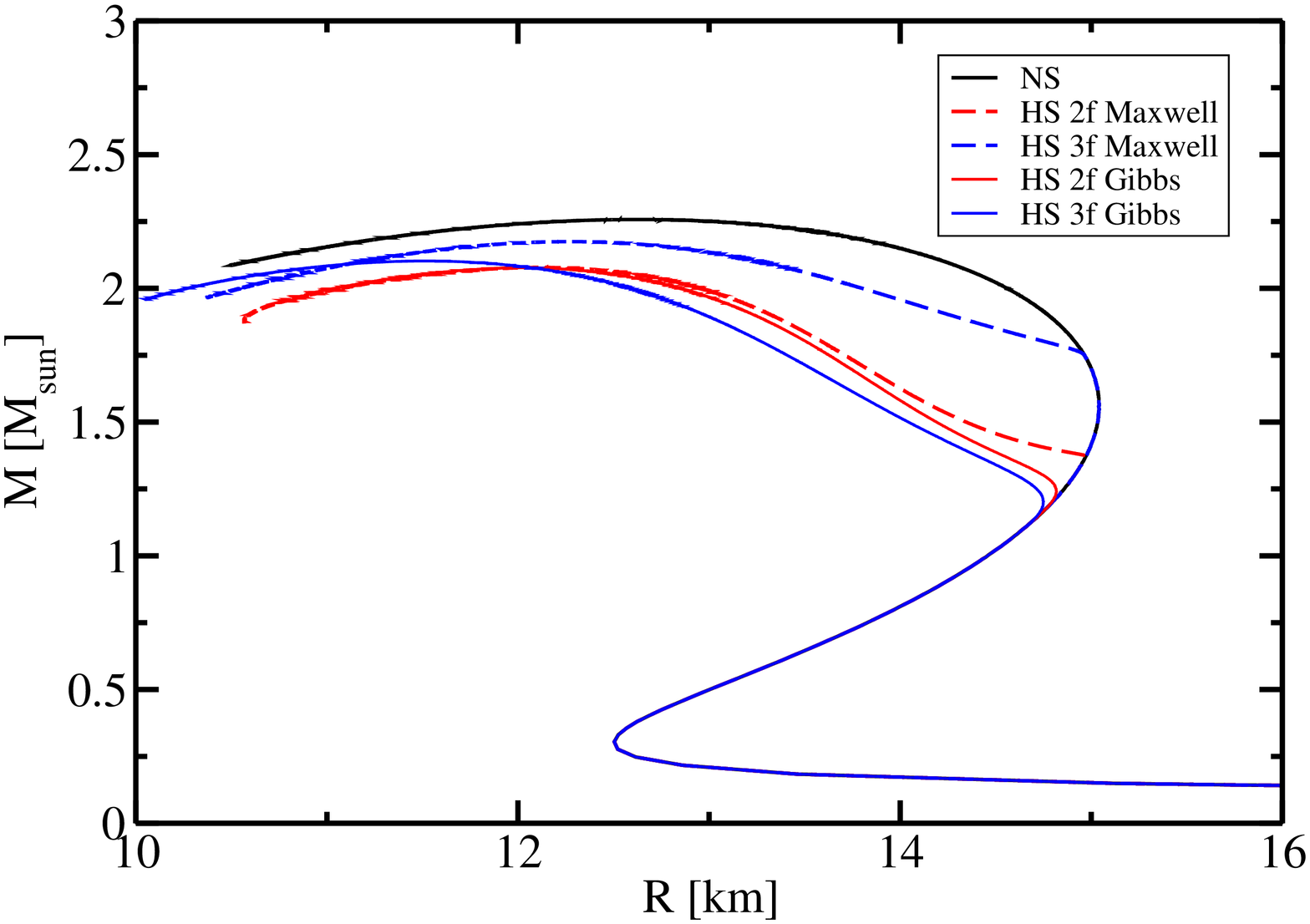}
\includegraphics[width=6.5cm,angle=270]{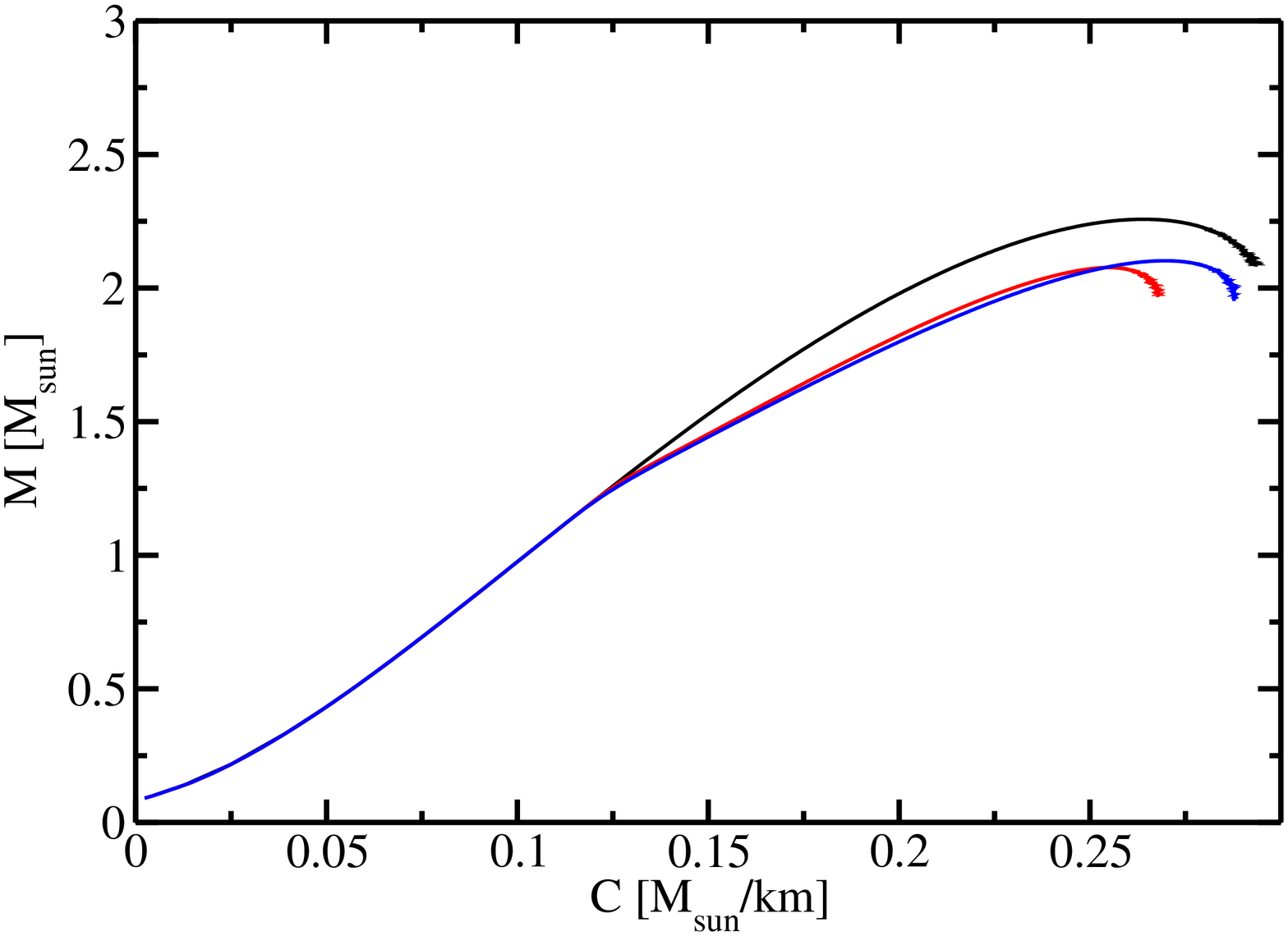}
\includegraphics[width=6.5cm,angle=270]{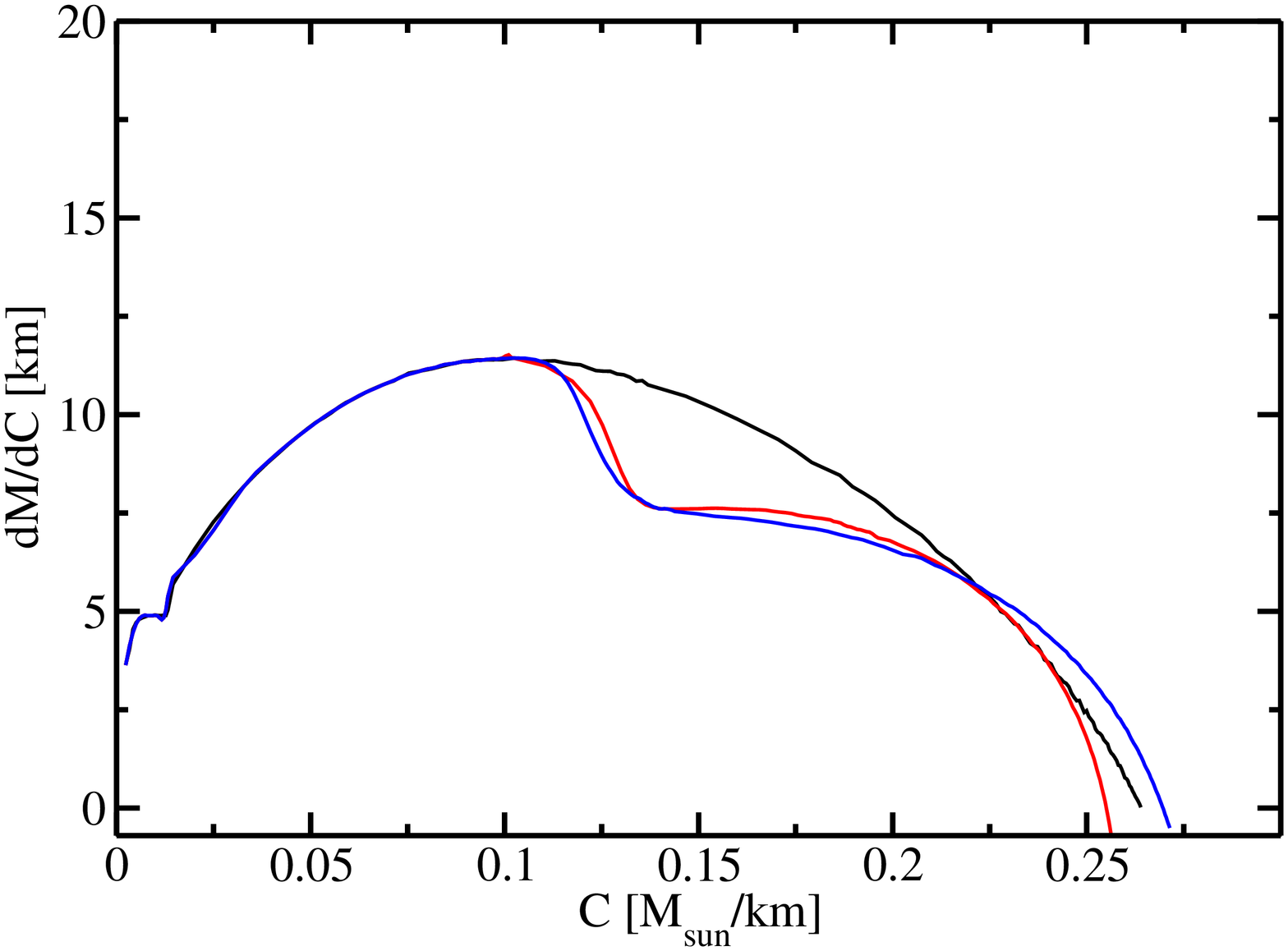}
\caption{Flavor camouflage for Gibbs type transitions: although different vBag parameterizations
are introduced, the transition to 2f quark matter
(solid red; $K_v$=$5\times10^{-6}\,{\rm MeV}^{-2}$, $B_{\rm eff}$=$70 {\rm \,MeV\,fm^{-3}}$)
and 3f quark matter
(solid blue; $K_v$=$22\times10^{-6}\,{\rm MeV}^{-2}$, $B_{\rm eff}$=$80 \,{\rm \,MeV\,fm^{-3}}$)
behave very similarly in mass, radius, compactness and even the compactness derivative d$M$/d$C$. For completeness, the Gibbs type transition is also plotted as dashed lines with identical color code.
}
    \label{fig:GibbsCamou}
\end{figure}

Uncertainties about the nature of the QCD phase transition can add more difficulties.
If the phase transition is of Maxwell type, it is accompanied by a typical kink in the
mass radius relationship. 
This kink might be subtle in extreme masquerade situations but it is a general feature.
For Gibbs type transitions, viz. assuming co-existing nuclear and quark matter phases,
the kink is smoothed out.

The same approach is followed in obtaining the results show in Fig.\ref{fig:GibbsCamou}, this time assuming a Gibbs type phase transition.
As before, all resulting curves for 2f and 3f matter are too similar to safely reject one of the scenarios in favor of the other.

\subsection{Nuclear Masquerade or Flavor Camouflage?}
In the previous cases we did {\it not} assume a sequential appearance of quark flavors.
Doing so now admits further possibility for ambiguities which we now illustrate.
\begin{figure}
    \centering
\includegraphics[width=6.5cm,angle=270]{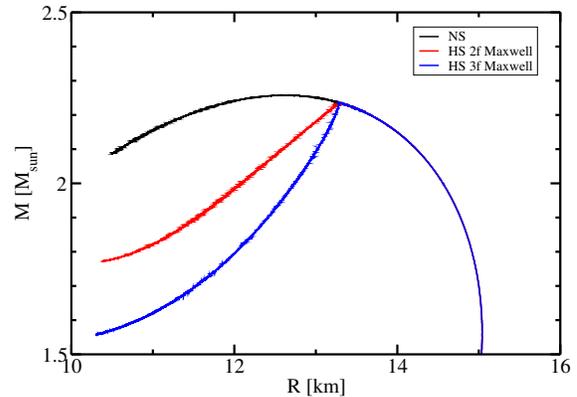}
\includegraphics[width=6.5cm,angle=270]{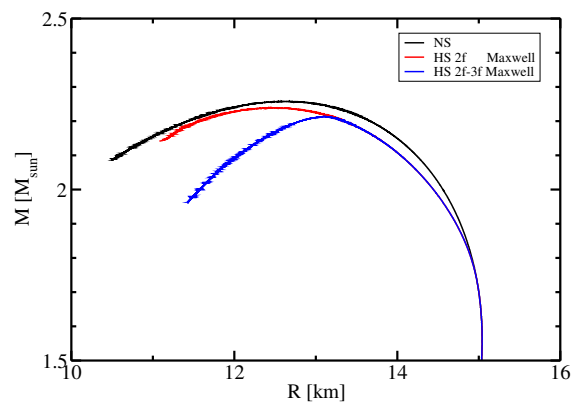}
\caption{A transition from nuclear to quark matter can result in unstable hybrid configurations, their onset marked by
a sudden cut of the $M$-$R$ relation 
(top panel; 2 flavor quark matter (red) with $K_v$=$1\times10^{-6}\,{\rm MeV}^{-2}$, $B_{\rm eff}$=$140\,{\rm MeV\,fm^{-3}}$
and 3 flavor quark matter (blue) with $K_v$=$1\times10^{-6}\,{\rm MeV}^{-2}$, $B_{\rm eff}$=$320\,{\rm MeV\,fm^{-3}}$).
A similar behavior can be observed for a sequential transition from a pure neutron star to a stable 2 flavor hybrid to an unstable 3 flavor hybrid
(bottom panel; 2 flavor quark matter (red) with $K_v$=$9\times10^{-6}{\rm MeV}^{-2}$, $B_{\rm eff}$=$70\, {\rm MeV\,fm^{-3}}$
and 3 flavor quark matter (blue) with $K_v$=$9\times10^{-6}{\rm MeV}^{-2}$, $B_{\rm eff}$=$270\,{\rm MeV\,fm^{-3}}$).
}
\label{fig:cutMR}
\end{figure}
A general feature of $M$-$R$ curves of purely nuclear EoS is a smooth approach
to the maximum mass with decreasing radius. 
A possible outcome of phase transition into quark matter is that the resulting hybrid stars
are unstable with masses below the maximum mass in the pure nuclear phase; the $M$-$R$ curve would end 'suddenly' before reaching
a maximum as described.
Although one could not observe an unstable hybrid star, the apparent 'cut' of the $M$-$R$ curve would
serve as an indicator that the phase transition is happening.
However, in accordance with the cases we discussed previously, one cannot safely conclude whether the transition went into 2f or 3f quark matter. 
This is illustrated in the top panel of Fig.\ref{fig:cutMR}.
Sequential chiral symmetry restoration, viz. the appearance of the strange quark at higher density than for the up and 
down quarks adds further subtlety.
We discussed earlier that for a sequential transition the appearance of the strange quark 
will always result in a Maxwell type appearance (due to the sudden increase of the energy density at a given pressure),
independent of which construction scheme one chooses to model the phase transition.
In the bottom panel of Fig.\ref{fig:cutMR} we show a scenario which would provide similar
data to the top panel case with a cut of the $M$-$R$ relationship before approaching a 'smooth' maximum.
However, in this scenario we chose to perform a Gibbs transition resulting in 2f masquerade already at masses
well below the maximum mass.
The transition from 2f to 3f quark matter results in unstable hybrid configurations and therefore in a cut of the $M$-$R$ plot
before approaching a 'smooth' maximum.
Starting from very different assumptions for the top (Maxwell transition from stable nuclear to unstable 2f or 3f quark matter) 
and bottom panel (Gibbs transition from stable nuclear to stable 2f mixed to unstable 3f mixed matter) 
we ended with $M$-$R$ relationships which seem indistinguishable even if high precision $M$-$R$ data were available.

\section{V. Conclusions}

The transition from nuclear to quark matter is not well understood and the lack of reliable first principle calculations
currently leaves us with only one option : to effectively model both independent equations of state and the nature of
the phase transition with the hope of benchmarking to available data and sort through the resulting parameter space to constrain and pursue further a few promising models. One of the key constraints, according to general opinion, is to come from measurements of neutron star radius with unprecedented precision in the very near future, narrowing the model space for the $M$-$R$ relationship. In this paper we have illustrated that our dependence on effective models unfortunately leaves us with a significant amount of ambiguity which might not be resolved even with high precision M-R data, at least if one is interested in phase transitions and the composition of dense matter deep inside the star. We have shown a selection of possibilities to explain different families of $M$-$R$ relationships with neutron stars of very different internal structure. In particular, {\it strange quark matter might camouflage as two-flavor quark matter}. Although this might not be a very optimistic perspective, we consider it as motivation to continue to increase our efforts at identifying complementary observational signatures for phase transitions in compact stars and to build on these to further develop and better constrain quark matter models. We applied vBag as an effective quark matter model which combines certain advantages of two standard approaches, the thermodynamic bag model and NJL-type models. It should be emphasized that we fully exploited the parametric freedom vBag provides as an effective model. Although more sophisticated models might not agree quantitatively with all the scenarios we discussed, we stress that the transition domain inside a neutron star is poorly understood and essentially unconstrained from a first principles perspective, so that any theoretical results should be interpreted and applied with caution. That being the case, we prefer vBag's flexibility to account for neutron star data in different ways to perform a critical assessment of what we could possibly learn from neutron star $M$-$R$ data.


\emph{Acknowledgments.}---W. W. is supported by the Natural Science Foundation
of China under Grant No.11547021 and China Scholarship Council. P.J. is supported by the U.S. NSF Grant No. PHY 1608959. We acknowledge helpful discussions with Fridolin Weber, Marc Salinas and Megan Barry.


\bibliographystyle{unsrt}
\bibliography{paper1}
\end{document}